# Evidence for a two-fold symmetric superconducting gap in a monolayer of $FeSe_{0.5}Te_{0.5}$ on a topological insulator

A. Kamlapure[1*], S. Manna[1], L. Cornils[1], T. Hänke[1], M. Bremholm[2], Ph. Hofmann[2], J. Wiebe[1†]

and R. Wiesendanger[1]

[1]*Department of Physics, University of Hamburg, Jungiusstrasse 11, D-20355 Hamburg, Germany*

[2]*Interdisciplinary Nanoscience Center iNANO, Aarhus University, Denmark*

**We present our investigations on the superconducting properties of monolayers of $FeSe_{0.5}Te_{0.5}$ grown on the 3D topological insulator $Bi_2Se_{1.2}Te_{1.8}$ using low temperature scanning tunneling spectroscopy (STS). While the morphology and the overall transition temperature resemble those of similarly doped bulk crystals, the spatially resolved spectroscopic data at 1.1K shows a much larger spatial inhomogeneity in the superconducting energy gaps. Despite the gap inhomogeneity all the spectra can be fitted with a two-fold anisotropic s-wave gap function. The two-fold nature of the gap symmetry is evident from the Bogoliubov quasiparticle interference (QPI) pattern which shows distinct $C_2$ symmetric scattering intensities. We argue that the gap inhomogeneity emerges as a result of intrinsic disorder in our system similar to disordered conventional superconductors. Even though most of our findings clearly differ from the current understanding of the corresponding bulk system, it provides an ideal platform to study unconventional superconductivity in Fe chalcogenides thinned down to a single layer and in close proximity to a topological insulator.**

[*]akamlapu@physnet.uni-hamburg.de
[†]jwiebe@physnet.uni-hamburg.de

The discovery of Fe based superconductors (FeSC) is an important hallmark in the field of superconductivity [1]. It provides the potential route to understand the microscopic mechanism of unconventional superconductivity in high $T_c$ cuprates because of the analogous phase diagram of both materials featuring an antiferromagnetically ordered parent compound. The electronic structure of FeSC is fairly complex with both electron and hole multibands. In these intrinsically multiorbital systems both spin fluctuations [2] and orbital fluctuations [3] have been argued to be responsible for the origin of superconductivity. However, recent investigations have shown that strong spin-orbit coupling lifts degeneracies between different $d$ bands and is intimately linked to the observed electronic anisotropy and nematicity [4][5]. As far as the stoichiometry is concerned, the Fe chalcogenides (FeCh), $Fe_{1+\delta}Se_xTe_{1-x}$, are the simplest systems of the different families of FeSC with optimal $T_c$ ~ 15.2K around x = 0.5 [6]. Regardless of the simple structure, the electronic and magnetic properties of FeCh are extremely sensitive to the growth conditions and pressures [7][8][9][10][11]. While the parent compound $Fe_{1+\delta}Te$ is not superconducting and exhibits bicollinear antiferromagnetism [12], superconductivity in FeTe is induced by tensile stress [7]. The growth delicacy is also observed for mixed FeCh where the $T_c$ of the films strongly depends on the substrates and is argued to be dependent on the ratio of the lattice parameters c/a [8][9]. Moreover, FeCh have recently gained interest due to the observation of record high $T_c$ in monolayers of Fe chalcogenides on $SrTiO_3$ substrates [13][14].

Concerning the order parameter (OP) for the bulk FeCh systems, numerous works suggest s± pairing [15][16]. However, there is a dispute about the s± picture [17][18] which makes the exact structure of the OP unclear. The V-shaped feature at low bias in the superconducting gap observed in MBE grown FeSe [19] supports nodal superconductivity while recent results on the nematic behavior of FeCh, resulting from the lifting of orbital degeneracy, support s++ pairing of the OP [18][20][21][5][22]. Also the energy gap in both bulk and thin films of FeSe is argued to be anisotropic with two-fold symmetry [19][23] while that in optimally doped bulk FeCh is shown to be four-fold symmetric [24][25]. All these findings on bulk and thin films of FeCh appear to be controversial and require in-depth studies of FeCh as a function of doping.

In this article we explore the superconducting properties of the monolayer of $FeSe_{0.5}Te_{0.5}$ grown on the topological insulator $Bi_2Se_{1.2}Te_{1.8}$ by high-resolution scanning tunneling spectroscopy (STS) which is a powerful tool to study electronic properties of FeCh in the nascent states. We observe inhomogeneous superconductivity in the monolayer of FeCh along with distinct $C_2$ symmetry in spectral images and we will discuss the inhomogeneity in the superconducting energy gap in the context of high $T_c$ cuprates and strongly disordered conventional superconductors. We also discuss our system in terms of its potential to study hybrid systems constituting a superconductor and a topological insulator.

In order to grow monolayers (ML) of $FeSe_xTe_{1-x}$, we evaporated nominally 0.5ML of Fe on the topological insulator $Bi_2Se_{1.2}Te_{1.8}$ [26] at room temperature under ultra-high vacuum conditions and subsequently annealed the sample at 300°C for about 15 minutes. Figure 1(a) shows 3D view of a large area STM topograph where we see the growth of large islands of monolayers and small islands of two layers of $FeSe_xTe_{1-x}$ [27]. The line profile across three islands is highlighted in Figure 1(b). The height of the monolayer above the substrate is ~ 0.71nm which is slightly larger than the corresponding bulk lattice constant (0.607nm) [28], while the height of the second layer with respect to the first layer is ~ 0.6 nm. Constant current topographs with atomic resolution taken on the substrate depict the hexagonal atomic structure as seen in Figure 1(c), while the STM topograph acquired at the center of one of the islands of $FeSe_xTe_{1-x}$ presented in Figure 1(d) shows the tetragonal atomic structure which resembles the bulk-like atomic structure reported earlier [29]. Here, Te appears brighter than Se atoms and by counting the number of each species based on the apparent heights we estimate the composition as ~ (50 ± 10)% Te and Se each, which is the optimal doping for highest $T_c$ in the corresponding bulk system. From the line profile (blue) shown in Figure 1(e) the height difference between Te and Se atoms is ~ (32±10) pm. From a Fourier analysis of the STM data on different islands we get the in-plane lattice constant ~ (0.38 ± 0.002) nm, which is close to that of the bulk material [28]. We also see a slight difference (~1-3%) in the lattice constants in the two directions; however, we cannot distinguish between the 'a' and 'b' directions. We acquired spectra of the electronic structure on the films by measuring the differential tunneling conductance (dI/dV) as a function of bias voltage with bulk Cr tip using lock-in technique [30]. Figure 1(f) shows the tunneling conductance spectra acquired along a line of 8 nm length starting with a Te rich site displaying a fully developed gap at the Fermi level ($E_F$) characteristic for superconductivity. It is interesting to note that on Te rich sites we observe two gap features with two coherence peaks that appear symmetrically around $E_F$ at (2.1 ± 0.5) mV and (4.5 ± 0.5) mV (Figure S1). The smaller gap observed in our measurements is consistent with earlier STS results [15][31]. However, the larger gap characteristic for the Te rich sites has not been reported earlier. Interestingly, the large gap value resembles the value reported in bulk ARPES measurements [32].

To study the spatial evolution of the superconductivity we acquired dI/dV spectra at T = 1.1K on 60 x 60 pixels over a 20nm x 20nm area. Figure 2(b-d) shows characteristic spectra which are symmetrized around zero bias and acquired at three different locations shown in the topographic images by corresponding circles. We observe an inhomogeneity in the superconducting spectra which is also visible in the tunneling conductance maps at different bias values [Figure S2]. To model our data we use BCS-Dynes theory [30] with an anisotropic energy gap given as $\Delta(\theta) = \Delta_0[1 + a(\cos 2\theta - 1)]$ and the Dynes broadening parameter $\Gamma$. Here $\Delta_0$ is the maximum value of the energy gap and $a$ represents the degree of gap anisotropy [Figure S3]. It should be noted that the choice of the anisotropic gap function is not unique here. [Figure S3]. The spectrum in Figure 2(b) is fitted with the parameters $\Delta_0 =$

1.85 meV; $a = 0.22$; $\Gamma = 0.17$ meV while those for the spectrum in Figure 2(d) are $\Delta_0 = 1$ meV; $a = 0.25$; $\Gamma = 0.15$ meV. The spectrum in Figure 2(e) is characteristic for Te rich sites as discussed earlier and can be fitted with a two gap model using the equation $G = \sigma G_L + (1-\sigma)G_S$, where $G_L(G_S)$ is the differential conductance simulated using a large (small) energy gap $\Delta_{0L}$ ($\Delta_{0S}$), $\sigma$ is the spectral weight due to the large gap and $G$ is the resulting conductance [33]. The fit parameters used are $\Delta_{0L} = 4.5$ meV; $\Delta_{0S} = 2.2$ meV; $a = 0.22$; $\Gamma = 0.15$ meV; $\sigma = 0.35$. It should be noted that here we used the same degree of anisotropy and lifetime broadening parameter for both the gaps. It is evident from these fits that we are able to capture the spectral shape within the anisotropic gap framework barring the high bias background with an inverted dome shape. To analyze the data acquired on the entire area we remove this background by dividing all the spectra with the spatially averaged spectrum taken at 12 K, where we see only the background with no signature of superconducting correlations, and fit them using an anisotropic gap function with an automatic fitting program which uses the $\chi^2$ minimization algorithm. For simplicity we have fitted the spectra only with a smaller gap and leave out the larger gap. The resulting energy gap values are plotted as a map in Figure 2(e) in the form of an intensity plot. We see quite a large variation in the gap values ranging from 0.45 meV to 3 meV and formation of patches with irregular shapes. Figure 2(f) shows the anisotropy map obtained from the corresponding fits to the spectra where $0 < a < 0.5$ represents an anisotropic and nodeless gap, while $a = 0.5$ corresponds to an anisotropic and nodal gap. It is clear from the anisotropy map that locally there is a quite large variation in the gap structure which changes from nodal to nodeless. In Figure 2(g) we plot the two-dimensional histogram of gap magnitude and differential tunneling conductance at V = 1mV ($g(\vec{r}, V = 1 \text{ mV})$). Here, we observe the expected anti-correlation between $\Delta$ and $g(\vec{r}, V = 1 \text{ mV})$ reflected by the negative slope, which we will use in our further analysis below. Similarly in Figure 2(h) we plot a two-dimensional histogram of the gap magnitude and the corresponding heights measured in the

STM topograph which displays an overall positive slope with large scatter. Although we find a very weak correlation between the topographic height and a gap magnitude, the larger inhomogeneity in the gap magnitude cannot be explained with our measurements alone. It is also important to take into account the role of substrate to modify the electronic structure of the film, where the charge doping from the substrate is known from the previous studies on similarly prepared FeSe films [34].

A similar gap inhomogeneity is well known from high $T_c$ cuprates [35] and disordered s-wave superconductors [36] where it is argued that the zero resistance state is lost through phase fluctuations [37]. It is therefore worth to compare our system with disordered superconductors. As far as local chemical inhomogeneity is concerned we see that Se rich and Te rich patches are uniformly spread all over the measured area and we do not see any indication for local strain in the film based on measuring local lattice constants from Fourier analysis. In our case, possibly the charge transfer between the substrate and film [34] strongly modifies the local density of states (DOS) owing to the complex nature of band structure of FeCh and conceivably act as a local disorder which affects the local pairing. In analogy to the behavior of disordered superconductors [36][38], large superconducting domains emerge in the system studied here over length scales of ~ 15-20 nm that are larger compared to the length scale for chemical phase separation (~ 1-2 nm). This scenario is further supported by the temperature dependent spectroscopy. In Figure 2(i) we show the temperature evolution of spectra acquired at a large gap location in different area along with the fits within the BCS-Dynes framework using an anisotropic gap as described above. The degree of anisotropy used for best fits at all temperatures is $a = 0.32$. Figure 2(j) depicts the temperature evolution of $\Delta$ and $\Gamma$. The evolution of $\Delta$ follows the solid black curve which is expected from BCS theory [39] with $T_c = 12.5$ K, while we observe that the comparatively large $\Gamma$ increases with temperature. The large $\Gamma$ in our system indicates short lived Bogoliubov quasiparticles, hence strong scattering seen in our system as discussed below. The intrinsic large scattering in our system is consistent with the similar observation in superfluid density measurements [40] in corresponding doped bulk FeCh. When we track the spatial evolution of dI/dV spectra as a function of temperature [Figure S4] we see that inhomogeneous spectra evolve smoothly and segregate spontaneously to form patches with large gap magnitude indicated by low zero bias conductance (blue) [Figure S4]. This clearly suggests a breakdown of long-range coherence at elevated temperatures through thermal phase fluctuations [37]. Also at low temperatures we cannot rule out (i) quantum phase fluctuations in our system as we have a single layer of FeSe$_x$Te$_{1-x}$ and (ii) the possible inhomogeneous charge transfer from the substrate over the length scale of the observed domain size [34]. It would be interesting to study this system under different growth conditions including a variation of the Se/Te concentration and different annealing times. So far our results motivate further theoretical studies to explore disordered superconductivity in Fe based superconductors.

Next, to get both real space and momentum space information on the electronic structure we acquired spectroscopy maps within an energy range of -5mV to 5mV over 512 x 512 pixels on a 25 nm x 25 nm area shown in Figure 3(a) (Figure S5). Figure 3(b) shows a tunneling conductance map at V = 1mV which again reveals a strong gap inhomogeneity owing to the anti-correlation between the conductance at V = 1mV and the energy gap (Δ) (See Fig. 2g). To minimize the set-point effects and enhance the visibility of the interference patterns of Bogoliubov quasiparticles we take the map of the ratio defined as $Z(\vec{r},V) = g(\vec{r},V)/g(\vec{r},-V)$ [41]. Figure 3(c) and Figure 3(e) show these ratio maps at V = 1mV and V = 2mV respectively. Here, we see a unidirectional stripe-like pattern, i.e. C$_2$ symmetry of the electronic structure, which is further evident in the corresponding autocorrelation maps in figure 3(d,f) where the periodic order corresponding to the distance of ~12a$_{Fe-Fe}$ is visible in a direction along Fe-Fe bond angle. Interestingly, similar patterns are also observed in Fe pnictides [42][43] where this observation has been associated with nematicity.

The profound $C_2$ symmetry of the electronic structure is also visible when we plot Fourier transform (FT) images of the ratio maps (Z maps) (Figure S6), where we observe quasiparticle interference (QPI) patterns being mainly distributed along and parallel to the Fe Bragg peaks, but not perpendicular. Figure 4(b-f) shows FT maps obtained from Z maps, where the Fe Bragg peaks are marked with white circles. To illustrate the dispersion of the different QPI scattering vectors we plot in Figure 4(g,h) the line cuts through the FT images of ratio maps (Z maps) at different bias voltages along the two directions from the center through the Fe Bragg peaks: Cut-1 is along the horizontal Fe Bragg peaks and Cut-2 is along the vertical direction. We see both dispersing and non-dispersing QPI scattering vectors **q** as a function of energy. Cut-1 shows two noticeable intensities: $q_1 \sim 0.5*q_{Fe-Fe}$ at V = 1mV which disperses and is probably the result of interband scattering of electron and hole bands shown in Figure 4(a). Second, $q_2 \sim 0.09*q_{Fe-Fe}$ which does not disperse and represents the same ordering ~ $12a_{Fe-Fe}$ as seen in the autocorrelation of the real space Z maps (Figure 3(d, f)). Cut-2 shown in Figure 2(h) also displays two major intensities: $q_1 \sim 0.5*q_{Fe-Fe}$ shows slight dispersion, while $q_3 \sim 0.21*q_{Fe-Fe}$ represents nondispersive intensities parallel to the horizontal Fe-Fe Bragg peaks, thereby strongly indicating the breaking of $C_4$ symmetry. We note here that the features corresponding to $q_3$ resemble those seen in the QPI maps of the parent compound of pnictides shown in reference [42]. It should also be noted that even though we observe inhomogeneity in the superconducting energy gap, a similar $C_2$ symmetry is observed locally in both the low gap and high gap region of Fig.3 (b) which implies two-fold anisotropy in the gap structure. Such a symmetry breaking could possibly arise from a slight strain in our film but within the limits of STM resolution we do not find any variation of the lattice constant.

Regarding the spectral properties and the local $T_c$ obtained from the temperature dependent data it appears that our system essentially exhibits comparable properties as the bulk $FeSe_xTe_{1-x}$ system. It is therefore interesting to compare our results with the results on bulk FeCh systems. It is known that for the corresponding bulk systems the strong spin-orbit interaction breaks the orbital degeneracy between the $d_{xz}$ and $d_{yz}$ orbitals [4][44] which leads to orbital splitting between the two corresponding bands [5]. Correspondingly, based on a STM study on $FeSe_{0.4}Te_{0.6}$ [22] it was proposed that orbital ordering is responsible for the symmetry breaking observed in the spectral maps. The authors estimated a splitting energy of 8mV by using the joint density of states approach in order to compare the band structure to the observed QPI. Using their result of the scattering vector as a function of the orbital splitting energy we estimate a splitting energy of 19.5mV in our case [Figure S7] which is comparable to the one reported by ARPES studies of bulk FeCh [5]. All these findings strongly suggest that the nematic order is competing with superconductivity in our thin films. Nevertheless, we also cannot rule out that orbital fluctuations provide the pairing glue supporting superconductivity in our system [3]. However, it would be interesting to study the electronic structure of our system at higher temperatures (T > $T_c$), where the superconductivity is suppressed.

In summary, we have successfully grown monolayers of Fe based superconductors on topological insulators and characterized the local electronic properties. We describe the spectra using an anisotropic energy gap and find inhomogeneities unlike in the corresponding bulk system. The gap magnitude very weakly correlates with the topography and we argue that the emergent nature of superconductivity is similar to disordered s-wave superconductors. We propose a two-fold anisotropic gap structure based on the observation of a pronounced $C_2$ symmetry in the QPI maps which probably is the result of a stronger spin-orbital coupling as in the bulk material. Finally, our sample system provides a platform to study superconductivity of FeCh in close proximity to a topological insulator, and it will be interesting to explore the expected Majorana physics [45].

**Acknowledgements**

We are indebted to Alexander Balatsky, Christopher Triola, Tim O. Wehling and Udai R. Singh for valuable discussions. This work has primarily been supported by the ERC Advanced Grant ASTONISH (No. 338802). L.C., Ph.H. and J.W. acknowledge partial support through the DFG priority program SPP1666 (grant No. WI 3097/2), and T.H. acknowledges support by the DFG under grant No. HA 6037/2-1. We thank the Aarhus University Research Foundation for supporting the bulk crystal growth.

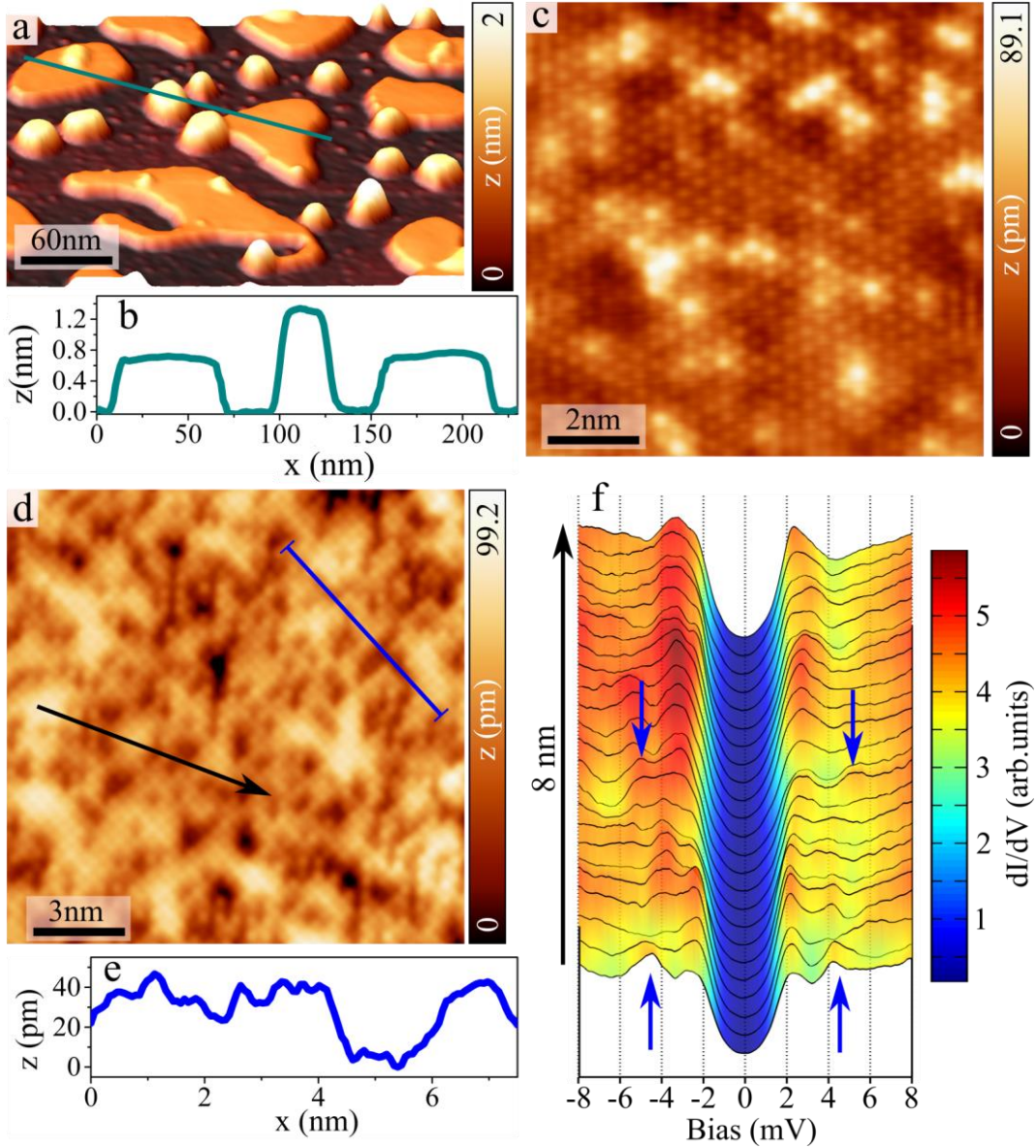

**FIG.1. Growth characterization of FeSe$_x$Te$_{1-x}$ on Bi$_2$Se$_{1.2}$Te$_{1.8}$**. (a) 3D view of a 300nm x 300nm constant-current STM topograph showing islands of FeSe$_x$Te$_{1-x}$ ($V$ = -600mV, $I_s$ = 30pA) (b) Height profile along the line shown in (a) revealing three islands. (c) Constant current-topograph acquired on the substrate ($V$ = -5.7mV, $I_s$ = 200pA). The image is a composition of the original and a Fourier filtered topograph to enhance the atomic resolution. (d) Constant-current topograph acquired on the single layer of FeSe$_x$Te$_{1-x}$. (x ~ 0.5). ($V$ = 200mV, $I_s$ = 3nA). (e) Height profile along the blue line shown in (d). (f) Tunneling spectra acquired at T = 1.1K at the position marked by the black arrow shown in (d). Blue arrows show the positions of second coherence peaks originating from a large gap structure characteristic for Te rich sites.

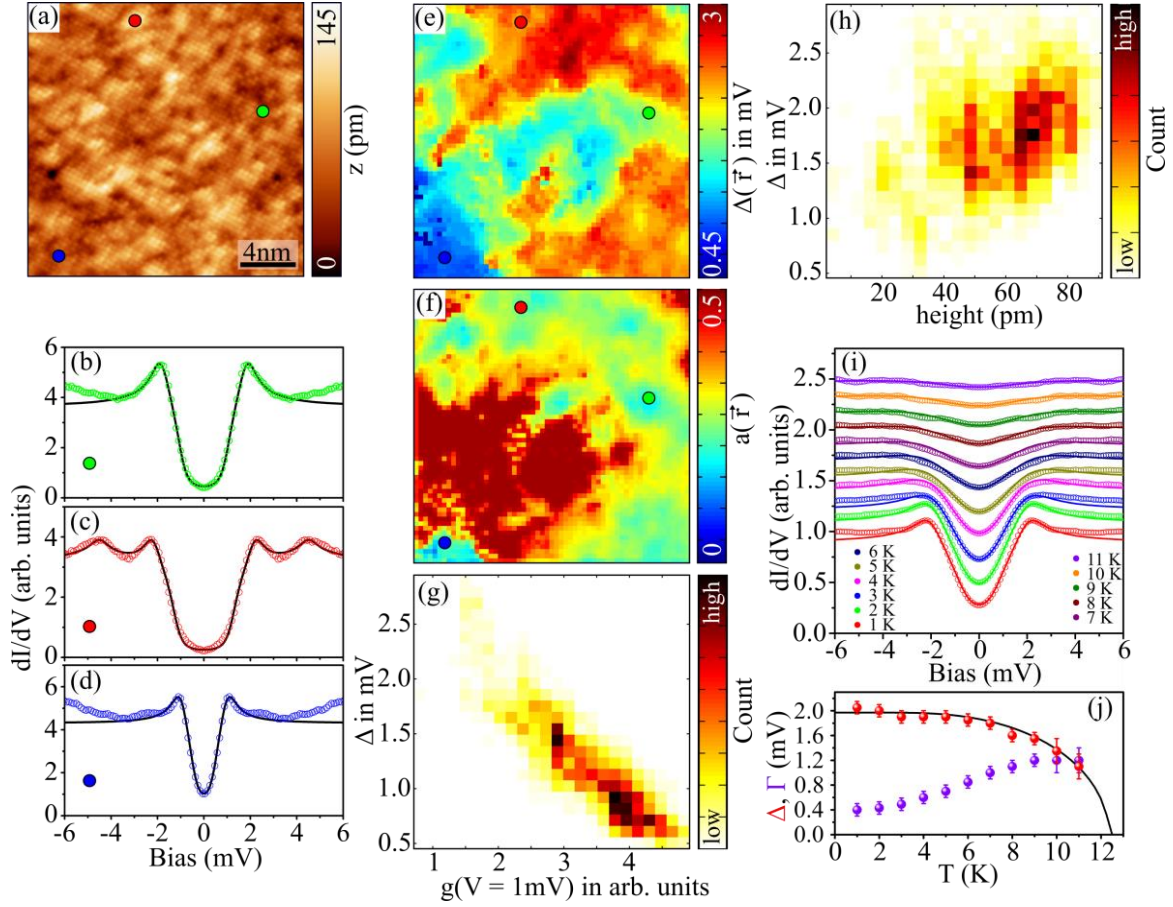

**FIG.2. Inhomogeneities of gap structure, gap size, and gap anisotropy.** (a) Constant current STM topograph showing an area of a single layer of $FeSe_xTe_{1-x}$ where the spectroscopic data of this figure has been measured ($V = 10mV$, $I_s = 400pA$). (b-d) Symmetrized tunneling spectra (markers) acquired at T=1.1K at the three different locations on $FeSe_{0.5}Te_{0.5}$ indicated by correspondingly colored markers in (a), (e), and (f). The continuous black curves plotted along with the measured spectra represent fits using an anisotropic gap function (see text). (e) Spatial evolution of the superconducting gap over the area of (a) obtained by fitting each symmetrized spectrum of a 60 x 60 pixel spectroscopic field employing an anisotropic gap in the BCS-DOS. (f) Corresponding anisotropy map obtained from the fits. (g) 2D histogram of the superconducting energy gap ($\Delta$) and the corresponding conductance value g at V=1mV plotted as intensity map. The negative slope here represents an anti-correlation between $\Delta$ and $g(V = 1mV)$. (h) 2D histogram of $\Delta$ and topographic height plotted as intensity map. Here, the positive slope with a large scatter represents a weak correlation between the two. (i) Symmetrized temperature dependent spectra along with anisotropic BCS fits. (j) Temperature dependence of the superconducting energy gap ($\Delta$) (red) and the Dynes broadening parameter ($\Gamma$) (purple). The solid black curve represents the temperature evolution of the energy gap $\Delta(T)$ expected within BCS theory.

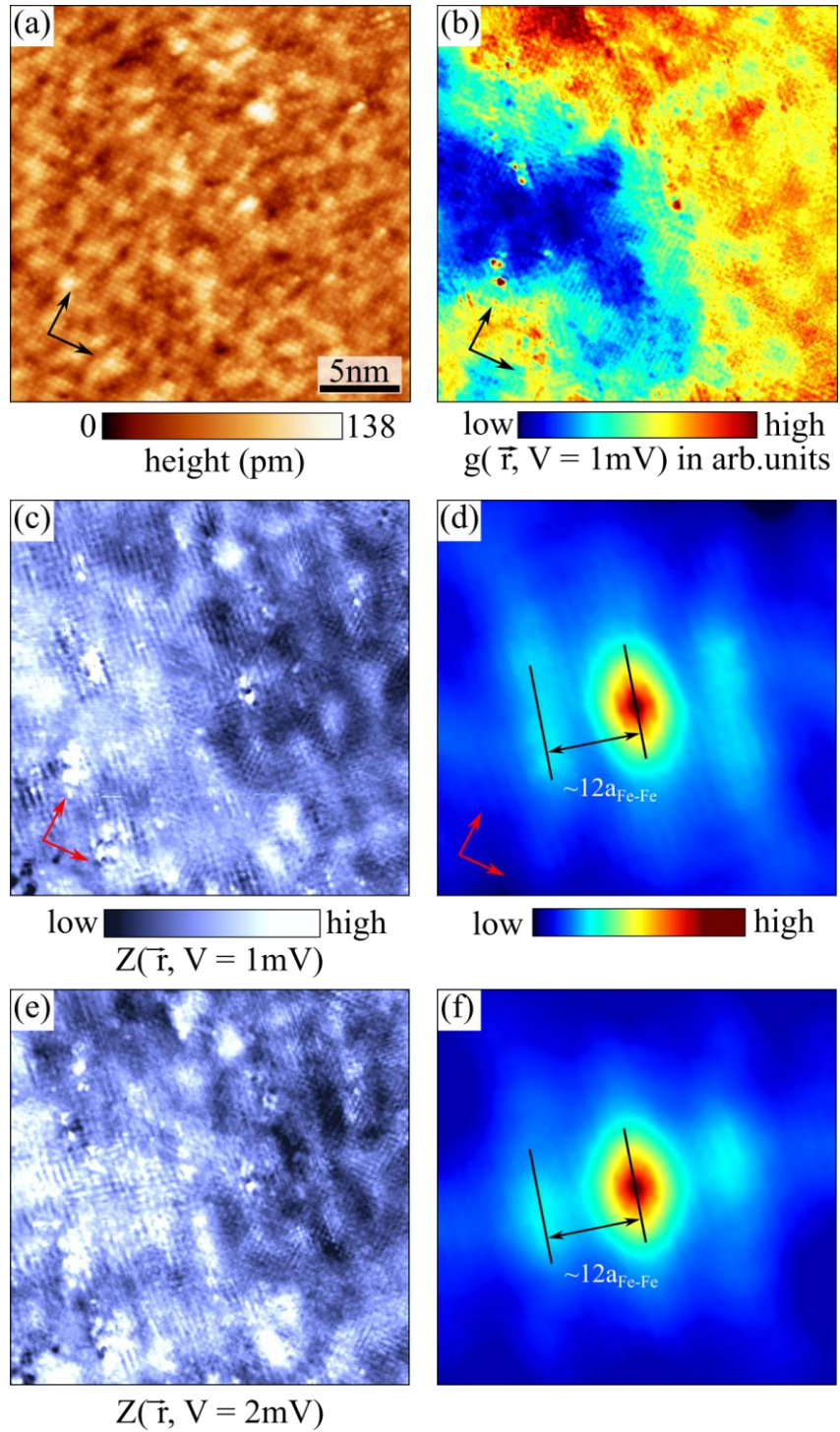

**FIG.3. Two-fold symmetry in autocorrelation of STS maps**. (a) Constant current STM topograph ($V = 10$mV, $I_s = 700$pA). (b) Conductance map at $V = 1$mV over the same area which represents the gap inhomogeneity. (c) Z map, obtained by taking the ratio of the conductance map at 1mV and -1 mV. (d) Autocorrelation map of Z map shown in (c). The arrows shown in the lower left side in each of the panels (a-d) represents the two lattice directions. (e) Z map at $V = 2$mV (f) Autocorrelation map of the Z map shown in (e). In the correlation plots the order corresponding to a distance of $12a_{Fe-Fe}$ is visible.

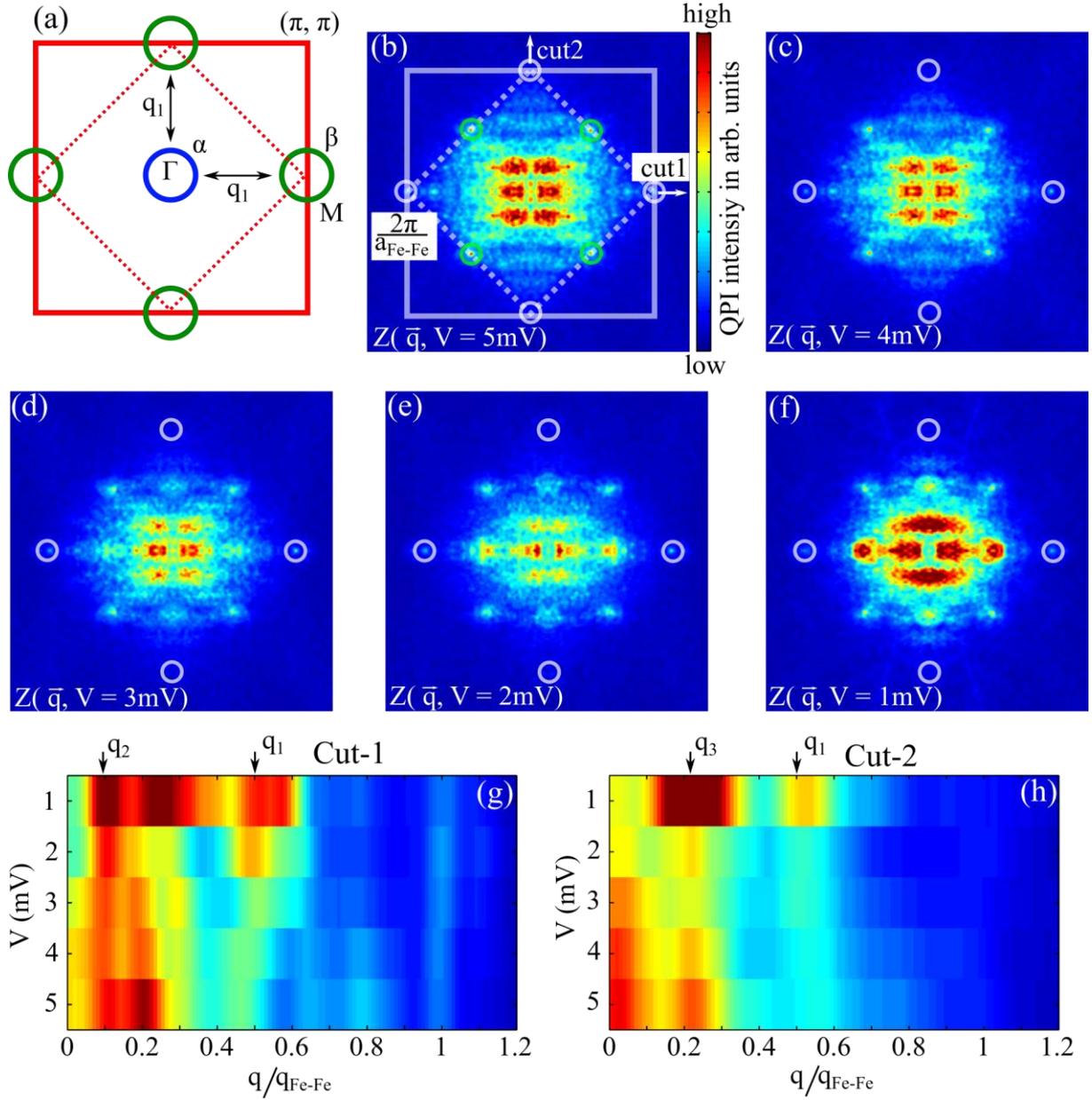

**Fig. 4. Two-fold symmetry in quasi-particle interference analyzed by FT-STS** (a) Simplified schematic diagram of the unfolded BZ of the unit cell with a single Fe atom. The blue circle at the gamma point represents a hole pocket and the green circles represent electron pockets. (b-f) Fourier transforms of Z maps from Fig.3 at the various indicated bias values. Light green circles in (b) represents Bragg peaks due to Se/Te atoms while white circles in each panel represent Bragg peaks due to Fe atoms. (g) Dispersion of QPI along Cut 1 shown in (b) which represents the high symmetry direction along Fe-Fe bonds. (h) Dispersion of QPI along Cut 2 shown in (b).

# Supplementary Material

## 1. Two gap features on Te rich sites

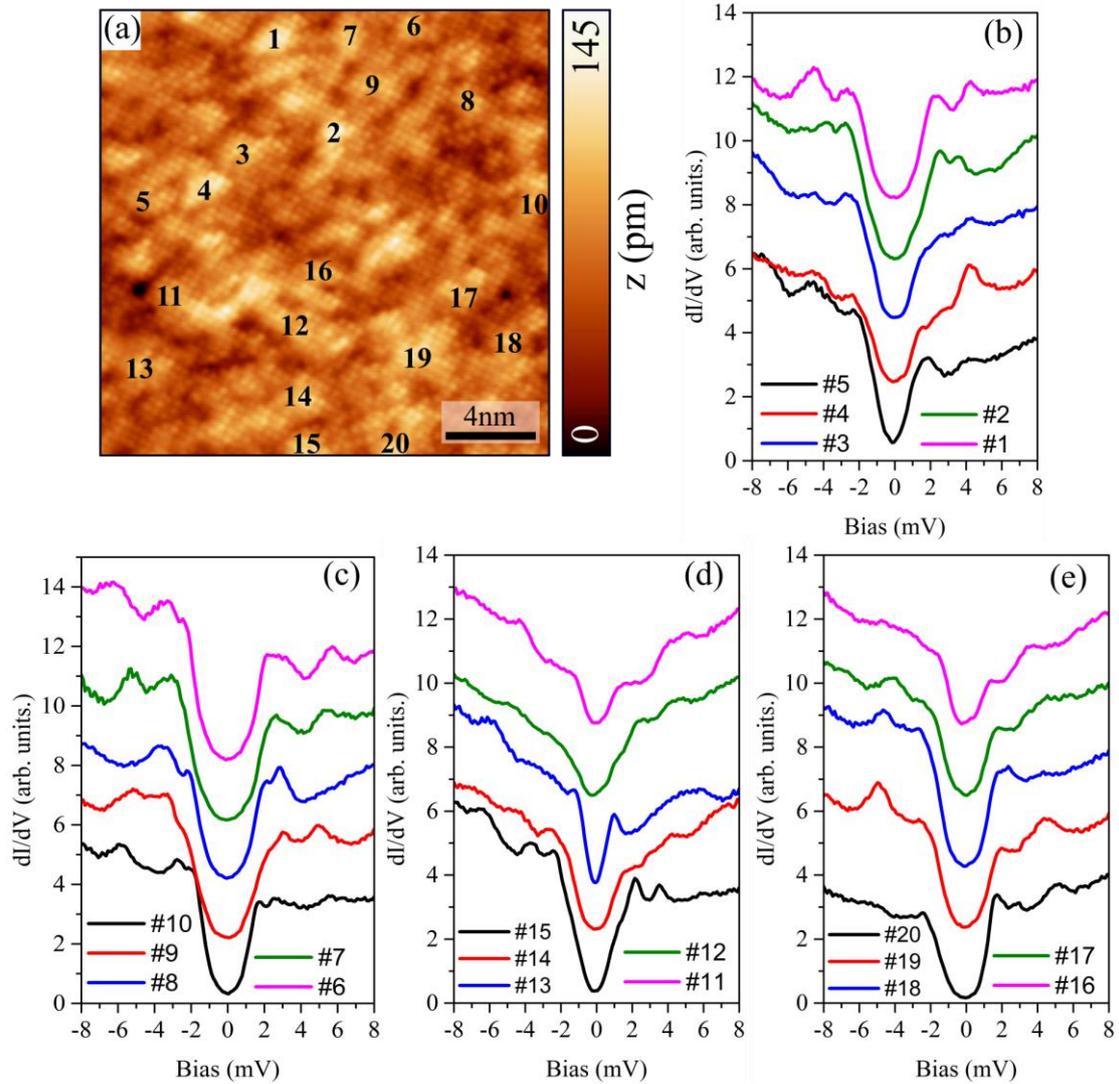

**FIG.S1.** (a) Constant current STM topograph acquired on single layer of FeSe$_x$Te$_{1-x}$ ($V$ = 10mV. $I_s$ = 400pA). Numbers on the topographic images indicate Te rich sites where spectroscopic data is acquired. The raw spectra corresponding to these numbers are plotted in (b-d). All the spectra show two gap features with two coherence peaks that appear symmetrical around $E_F$ at (2.1 ± 0.5) mV and (4.5 ± 0.5) mV.

## 2. Tunneling conductance maps

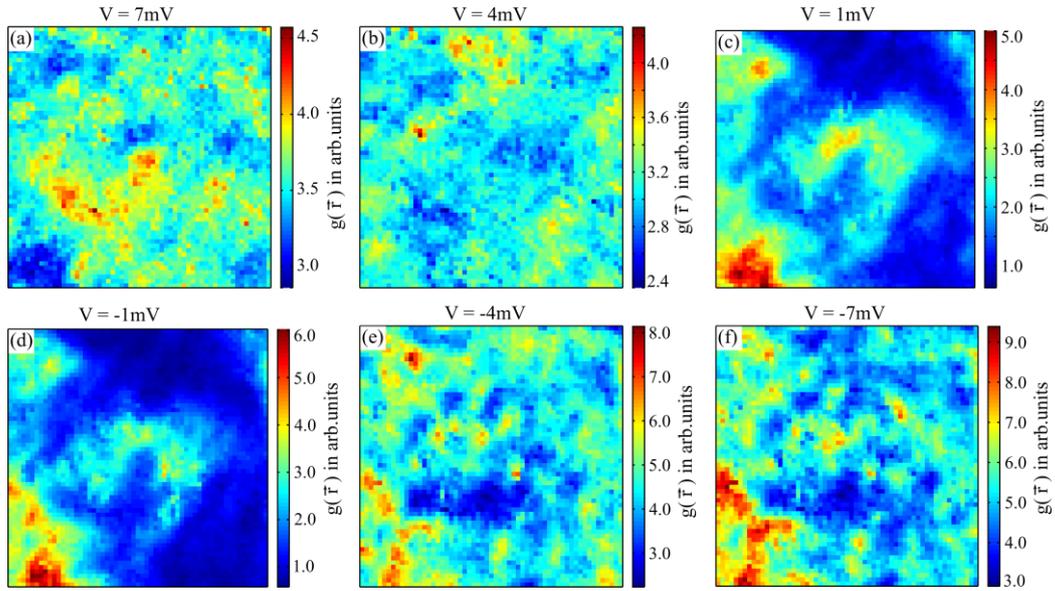

**FIG. S2.** Tunneling conductance maps at various bias voltages which are derived from the dI/dV spectra taken on 60 x 60 pixels over a 20nm x 20nm area. The inhomogeneity in the superconducting spectra is apparent from the maps.

## 3. Modeling STS data using an anisotropic energy gap

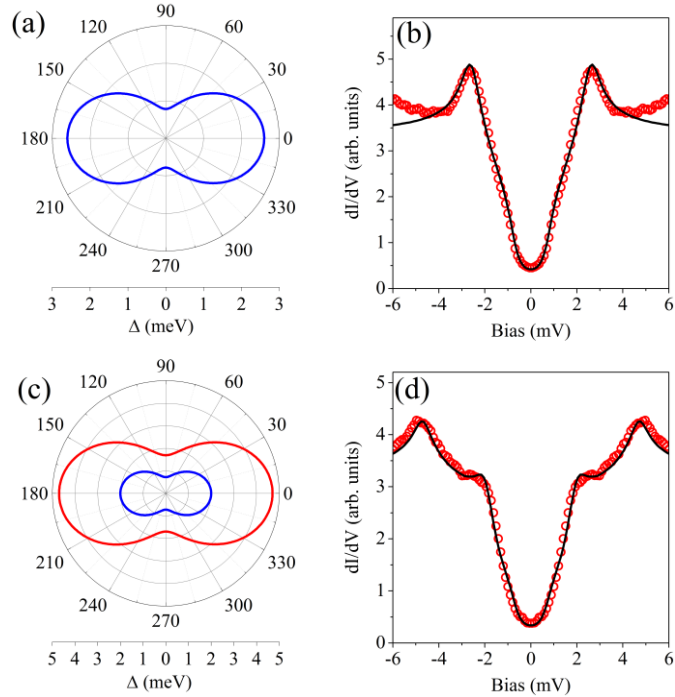

**FIG S3.** (a) Anisotropic gap function given by $\Delta(\theta) = \Delta_0[1 + a(\cos 2\theta - 1)]$ where $\Delta_0$ is the maximum value of the energy gap and $a$ represents the degree of gap anisotropy. In this case $\Delta_0 = 2.6\text{meV}$ and $a = 0.35$. (b) Typical symmetrized dI/dV spectrum (red circles) along with the black curve which represents a fit using the BCS-Dynes density of states $N(E) =$

$N_n(E_F) \cdot Re\left[\frac{E+i\Gamma}{\sqrt{(E+i\Gamma)^2-\Delta(\theta)^2}}\right]$ [1] employing an anisotropic energy gap $\Delta(\theta)$ shown in (a) and the Dynes broadening parameter $\Gamma = 0.17$meV. (c) Anisotropic gap function for two gaps with $\Delta_{0L} = 4.7$meV, $\Delta_{0S} = 2$meV and $a = 0.32$, where the subscripts $L$ and $S$ stand for large and small, respectively. (d) Typical dI/dV spectrum (red circles) taken at a Te rich site. The black curve plotted together with the spectrum represents a fit using a two gap model based on the equation $G = \sigma G_L + (1-\sigma) G_S$ [2], where $G_L(G_S)$ is the differential conductance simulated using large (small) energy gap $\Delta_{0L}$ ($\Delta_{0S}$), $\sigma$ is the spectral weight due to the large gap and $G$ is the resulting conductance. The anisotropic energy gaps used are shown in (c) while other fit parameters are: $\sigma = 0.5$ and $\Gamma = 0.17$meV. It should be noted that the choice of the anisotropic gap function is not unique. We get the same results if we replace $\cos 2\theta$ term in the gap function by $\cos 4\theta$.

## 4. Temperature evolution of tunneling conductance spectra

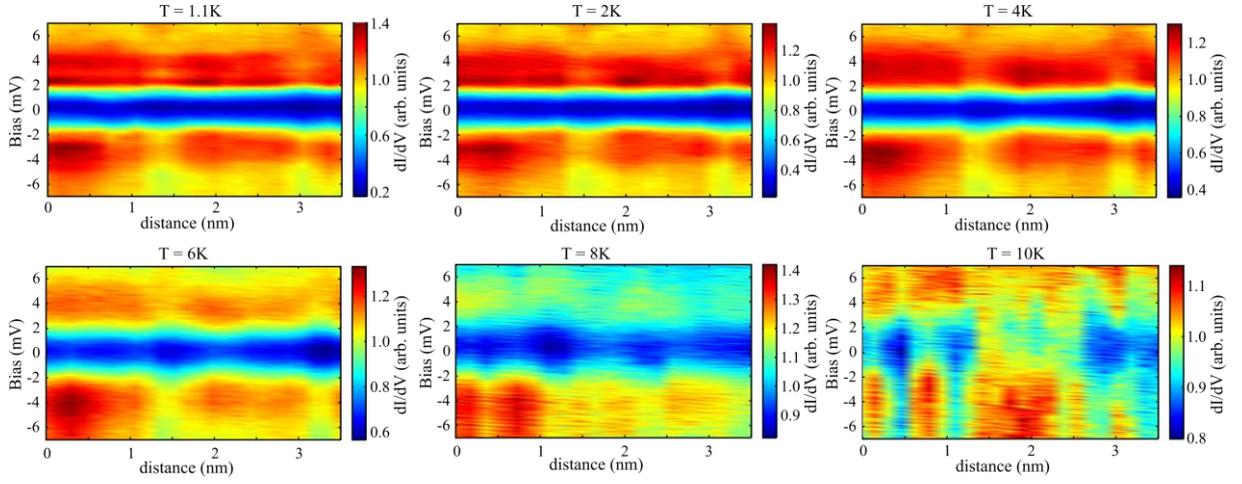

**FIG. S4.** Each panel here is a 2D plot of tunneling spectra as a function of applied bias and distance where the data is acquired on the same line. We observe that the spectra evolve smoothly and at high temperature, the local superconducting correlations persists as seen by the formation of blue patches related to a dip in dI/dV due to superconductivity.

## 5. Tunneling conductance maps corresponding to QPI data

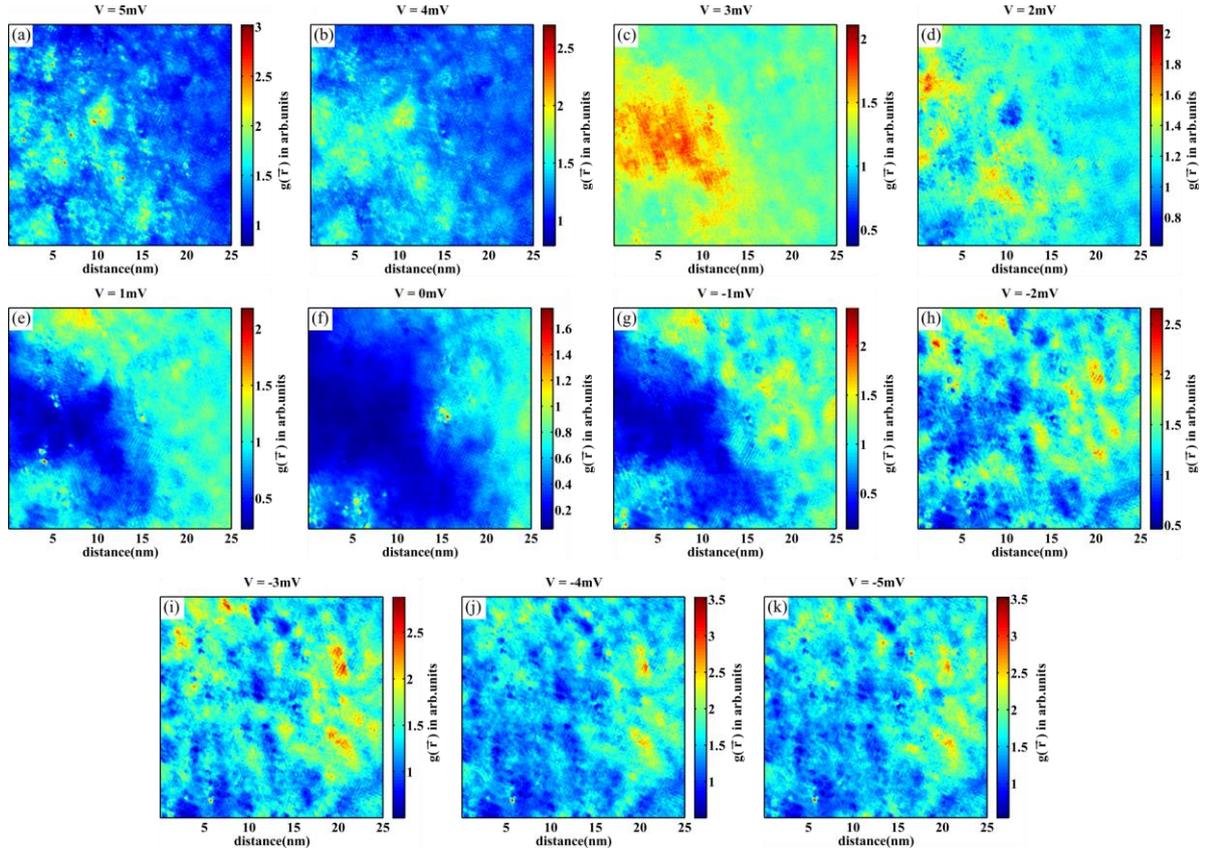

**FIG. S5.** Tunneling conductance maps at various bias voltages which are derived from the dI/dV spectra taken on 512 x 512 pixels over a 25nm x 25nm area. The inhomogeneity in the superconducting properties is apparent from these maps at low bias voltages

## 6. Fourier transform analysis

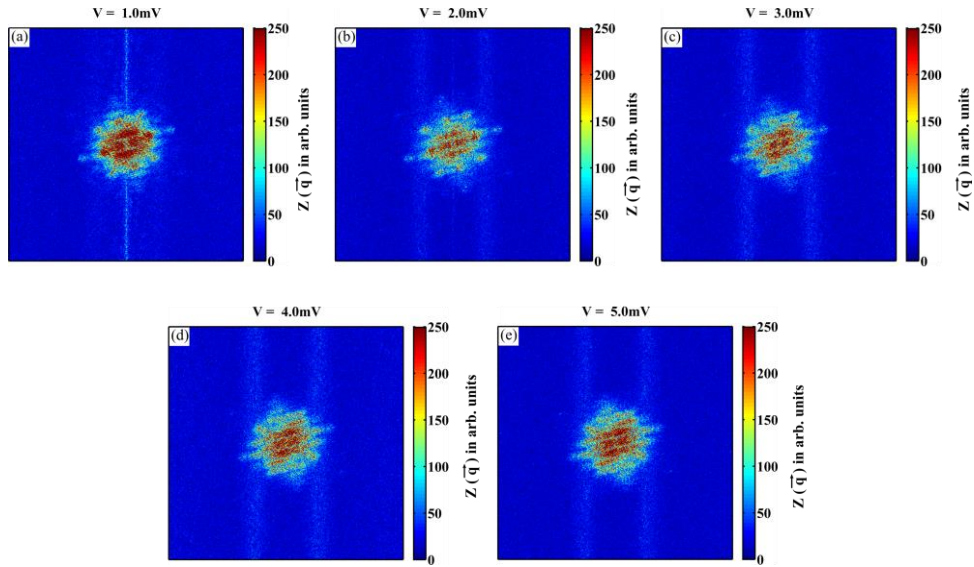

**FIG. S6.** Fourier transforms of Z maps at various bias values. Here each plot is obtained using fast Fourier transform analysis of Z maps and then subtracting the central core to enhance the visibility of relevant **q** vectors. For further analysis of different quasiparticle structures, the data is symmetrized along a high symmetry axis, i.e. along Fe-Fe bonds and low pass filtered to reduce the noise. The resulting data is presented in **FIG. 4.**

## 7. Estimation of orbital splitting energy

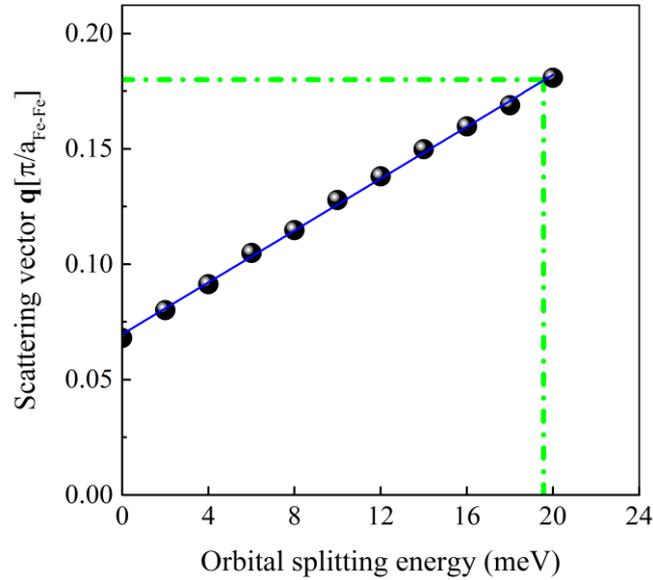

**FIG. S7. Estimation of orbital splitting energy.** The plots represent the data extracted from reference [3] where the authors calculated the orbital splitting energy as a function of the dominant scattering vector **q**, corresponding to symmetry breaking states, employing the joint density of states approach. Using their results we estimate the orbital splitting energy of 19.5meV in our system which corresponds to the observed scattering vector **q** = 0.18.